\documentstyle[aps,psfig]{revtex}
\begin{document}
\wideabs{
\title{Critical behavior of ${\bf YBa_{2}Cu_{3}O_{7-\delta}}$ and 
${\bf Bi_{2}Sr_{2}CaCu_{2}O_{8+\delta}}$ from the dual Ginzburg-Landau model}
\author{Claude de Calan and Flavio S. Nogueira}
\address{Centre de Physique Th\'eorique,  
Ecole Polytechnique, 
F-91128 Palaiseau Cedex, FRANCE}
\date{Received \today}
\tighten
\maketitle

\begin{abstract}
An unifying scenario for the critical behavior of
YBa$_{2}$Cu$_{3}$O$_{7-\delta}$ (YBCO) and 
Bi$_{2}$Sr$_{2}$CaCu$_{2}$O$_{8+\delta}$ (BSCCO) is proposed. It is 
shown that both critical crossovers observed in these materials follow 
by considering two different scalings in the dual Ginzburg-Landau model.  
The first scaling leads 
to the critical exponents $\nu\approx 2/3$, $\nu'\approx 1/3$ and 
$\alpha\approx 0$, with $\nu$ and $\alpha$ being respectively the 
correlation lenght and specific heat exponents while $\nu'$ is the 
magnetic field penetration depth exponent. These values for the 
critical exponents agree with the ones obtained experimentally for 
YBCO single crystals. For the second scaling we obtain $\nu=1$ and 
$\alpha=-1$ which must be compared with the measured values 
$\nu\approx 1$ and $\alpha\approx-0.7$ for BSCCO. For the penetration 
depth exponent it is predicted the value $\nu'=1$.      
\end{abstract}
\draft
}

It is generally accepted that the experimentally accessible critical region of 
high temperature superconductor    
YBa$_{2}$Cu$_{3}$O$_{7-\delta}$ (YBCO) lies in a 
$^4$He universality class \cite{Salamon,Kamal,Junod}. This means that 
the critical behavior is governed by the 3D XY ($XY_{3}$ for short)
nontrivial fixed point. The $XY_{3}$ critical behavior of YBCO has been 
probed both in zero and nonzero external field regimes \cite{Junod}. For 
instance, a quantity which can be accurately measured in zero field is 
the magnetic field penetration depth, $\lambda$, whose scaling behavior 
near $T_{c}$ is given by $\lambda\sim|t|^{\nu'}$, with 
$\nu'=0.33\pm 0.1$ \cite{Kamal} and $t$ being the reduced temperature. 
This value of $\nu'$ is consistent with a critical regime described 
by a $XY_{3}$ universality class where we expect $\nu'=\nu/2$ with 
$\nu\approx 2/3$ \cite{Fisher}. Also, specific heat measurements give 
$\alpha\approx -0.013$ \cite{Salamon,Junod}, which is also consistent 
with the $XY_{3}$ universality class. 
These results hold for single crystals of  
optimally doped YBCO. However, this scenario may not hold for 
thin films of YBCO. For example, the value $\nu'=1/2$ 
has been measured at zero field regime \cite{Lin,Paget} which may 
be associated either to a mean-field behavior \cite{Paget} or 
to some other critical crossover \cite{Kiometzis1,deCalan1}. It is worth 
to cite also the recent results of Charalambous {\it et al.} \cite{Char} 
for bulk YBCO. They reported different values of $\alpha$ 
depending on whether $T_{c}$ is approached from below or from above. 

The $XY_{3}$ universality class is not always shared by other cuprate 
superconductors. For instance, a different critical behavior is probed 
in Bi$_{2}$Sr$_{2}$CaCu$_{2}$O$_{8+\delta}$ (BSCCO). It is obtained that 
$\alpha\approx-0.7$ and $\nu\approx 1$ \cite{Junod1,Roulin}. 
This is not very different from 
the critical exponents of the 3D $O(N)$ model in the large $N$ limit 
where we find exactly $\alpha=-1$ and $\nu=1$ \cite{ZJ}. 
In the context of superconductors, these exponents are obtained in a 
Hartree approximation in external field but with the gauge field 
fluctuations being neglected \cite{Lawrie}. Indeed, the Hartree approximation 
consists of a gap equation analogous to the one obtained in the large $N$ 
limit of the $O(N)$ model. 
The critical exponents $\alpha=-1$ and $\nu=1$ can be obtained also through a 
weakly interacting Bose gas model \cite{Alex}. In this case the specific 
heat exhibits a triangular peak at zero field characteristic of a 
Bose-Einstein condensation (BEC). 

In this paper we will show that the critical crossovers observed in 
both YBCO and BSCCO are described by two 
different scalings in the dual version of the 
Ginzburg-Landau model \cite{Kiometzis1,Herbut1,Tesanovic} (to be referred 
from now on as the dGL model). 
The description of the $XY_{3}$ scaling of YBCO by the dGL model is 
well known \cite{Herbut1,deCalan1}. However, we want to emphasize the fact 
that two different crossovers are obtained here from a {\it same} model. 
We provide in this way an unifying view for the YBCO and BSCCO critical 
behaviors. 

The dGL model was introduced in the literature as a continuum 
version of the lattice dual Ginzburg-Landau (GL) model \cite{Kiometzis2}. The 
dGL model is therefore a way towards a field theoretical description of 
the inverted $XY_{3}$ transition \cite{Dasgupta}. The inverted $XY_{3}$ 
scenario must be valid at least for superconductors in the type II 
regime and ensures the existence of a {\it charged} infrared stable fixed 
point for the GL model \cite{Folk,Berg,Herbut2,deCalan2}. One important 
difference of the inverted $XY_{3}$ relative to the 
ordinary $XY_{3}$ universality class is the scaling 
$\lambda\sim\xi$ \cite{deCalan1,Herbut2,Olsson}, 
where $\xi$ is the correlation lenght. This scaling implies that 
$\nu'=\nu$ instead of $\nu'=\nu/2$ as in the ordinary $XY_{3}$ fixed point. 
Recent high precision Montecarlo simulations give further support to the 
duality scenario and to the existence of the inverted $XY_{3}$ critical 
point \cite{Sudbo,Nguyen,Hove}. 
Unfortunately, the critical region corresponding 
to this nontrivial charged fixed point is still experimentally out of 
reach.

The bare free energy density for the dGL model is given by 

\begin{eqnarray}
\label{dGL}
\cal{F}&=&\frac{1}{2}[(\nabla\times{\bf h}_{0})^2+M_{0}^2{\bf h}_{0}^2]
+|(\nabla-ie_{0,d}{\bf h}_{0})\psi_{0}|^2\nonumber\\
&+&m_{0}^2|\psi_{0}|^2+\frac{u_{0}}{2}|\psi_{0}|^4, 
\end{eqnarray}
where the constraint $\nabla\cdot{\bf h}_{0}=0$ should be understood. 
The bare dual charge $e_{0,d}=2\pi M_{0}/q_{0}$ where $q_{0}=2e_{0}$ is the 
Cooper pair charge. Also, $M_{0}^2=q_{0}^2\langle|\phi_{0}|^2\langle$ 
where $\phi_{0}$ is the bare order parameter field of the GL model. Thus, 
$M_{0}$ is the photon mass generated in the Meissner phase of the GL model. 
The Meissner phase of the GL model is described as 
a symmetric regime in the dGL model. In this sense, the dGL model is 
a disorder field theory. 

Due to the presence of two masses in the problem, we have  
more possibilities of scalings in the dGL model than in the GL model. A 
class of such scalings was discussed recently by us \cite{deCalan1}. 

The scaling corresponding to the $XY_{3}$ behavior of YBCO is obtained 
by looking to the flow of the renormalized dimensionless couplings 
$f=e_{d}^2/m$ and $g=u/m$ with respect to $m$ at $M_{0}$, $u_{0}$ and 
$e_{0,d}$ {\it fixed}. The renormalized parameters are defined 
in a usual way \cite{ZJ}. They are given by $m^{2}=Z_{m}^{-1}Z_{\psi}m_{0}^2$, 
$M^2=Z_{h}M_{0}^2$, $e_{d}^2=Z_{h}e_{0,d}^2$ and $u=Z_{\psi}^2 u_{0}/Z_{u}$. 
Such a scaling has been considered before in Refs. \cite{Herbut1,Nogueira} 
and for this reason we will omit the details. In this scaling the 
infrared stable fixed point corresponds to a nonzero $f$. As a 
consequence, the anomalous dimension 
$\eta_{h}=m\partial\ln Z_{h}/\partial m$ is given at the nontrivial 
fixed point by $\eta_{h}=1$ {\it exactly} \cite{Herbut1,Nogueira}. Since 
$M_{0}$ is fixed it follows the scaling $M\sim m^{1/2}$ near the critical 
point. Thus, it follows that $\nu'=\nu/2$ and the ratio $m/M\to 0$ as 
$m\to 0$. From this last observation we obtain that the gauge degrees 
of freedom decouple and the corresponding fixed point is $XY_{3}$. We 
have therefore that $\nu\approx 2/3$, $\nu'\approx 1/3$ and 
$\alpha\approx 0$, which are just the measured values in the bulk YBCO with 
optimal doping \cite{Salamon,Kamal,Junod}. 

For the scaling corresponding to the BSCCO critical behavior it is more 
convenient to choose $M$ as the running scaling variable. The BSCCO scaling 
is defined by demanding that the ratio $m_{0}/M_{0}$ is kept fixed together 
with $e_{0,d}$ and $u_{0}$ \cite{Note}. 
We assume therefore that both $m_{0}^2$ and $M_{0}^2$ are 
proportional to the reduced temperature $t$. In this scaling the 
renormalized quantities are defined as before and the corresponding 
renormalization constants are denoted by $\tilde{Z}_{\psi}$, 
$\tilde{Z}_{h}$, $\tilde{Z}_{m}$ 
and $\tilde{Z}_{u}$. The dimensionless couplings are now 
defined by $\tilde{f}=\tilde{Z}_{h}e_{0,d}^2/M$ and 
$\tilde{g}=\tilde{Z}_{u}^{-1}\tilde{Z}_{\psi}^2 u_{0}/M$. 
It is convenient to define also 
the dimensionless parameter $\kappa_{d}=m/M$ which plays a role analogous 
to the Ginzburg parameter. Note that fixing $e_{0,d}$ is equivalent to 
consider that the amplitude fluctuations are frozen in the original 
GL model. We define the renormalization group (RG) 
functions: 

\begin{equation}
\tilde{\eta}_{\psi}=M\frac{\partial\ln\tilde{Z}_{\psi}}{
\partial M},
\end{equation}

\begin{equation}
\tilde{\eta}_{h}=M\frac{\partial\ln\tilde{Z}_{h}}{
\partial M},
\end{equation}

\begin{equation}
\tilde{\eta}_{m}=M\frac{\partial\ln\tilde{Z}_{m}}{
\partial M}.
\end{equation}
We have the following {\it exact} flow equations: 

\begin{equation}
\label{f}
M\frac{\partial\tilde{f}}{\partial M}=(\tilde{\eta}_{h}-1)\tilde{f},
\end{equation}

\begin{equation}
\label{kappa}
M\frac{\partial\kappa_{d}}{\partial M}=\frac{1}{2}(\tilde{\eta}_{\psi}
-\tilde{\eta}_{m}-\tilde{\eta}_{h})\kappa_{d},
\end{equation}

\begin{equation}
\label{m}
M\frac{\partial m^2}{\partial M}=(2+\tilde{\eta}_{\psi}-\tilde{\eta}_{m}
-\tilde{\eta}_{h})m^2,
\end{equation}

\begin{equation}
\label{M0}
M\frac{\partial M_{0}^2}{\partial M}=(2-\tilde{\eta}_{h})M_{0}^2.
\end{equation}
A nontrivial fixed point with 
$\tilde{f}$, $\tilde{g}$ and $\kappa_{d}$ nonzero must  
verify the equations $\tilde{\eta}_{h}=1$,  
$\tilde{\eta}_{\psi}-\tilde{\eta}_{m}=1$ and 
$M\partial\tilde{g}/\partial M=0$.  
If such a fixed point exists, 
we obtain from Eqs. (\ref{m}) and (\ref{M0}) that $\nu'=\nu=1$ and 
$\alpha=-1$ from the scaling relation $3\nu=2-\alpha$, which are 
the required exponents. Note that by assuming the existence of the 
nontrivial fixed point we have that the values of the exponents are 
{\it exact} and are the same as in the BEC transition. 
The existence of this nontrivial fixed point can be 
verified approximately through a simple 1-loop example. We have at 
1-loop order,  

\begin{equation}
\tilde{\eta}_{\psi}=-\frac{2}{3\pi}\frac{\tilde{f}}{1+\kappa_{d}},
\end{equation}

\begin{equation}
\tilde{\eta}_{h}=\frac{\tilde{f}}{24\pi\kappa_{d}},
\end{equation}

\begin{equation}
\tilde{\eta}_{m}=-\frac{\tilde{g}}{4\pi\kappa_{d}},
\end{equation}

\begin{equation}
M\frac{\partial\tilde{g}}{\partial M}=(2\tilde{\eta}_{\psi}-1)\tilde{g}
+\frac{5\tilde{g}^2}{8\pi}+\frac{\tilde{f}^2}{2\pi}, 
\end{equation}
where in neglecting higher order terms we have assumed 
$\kappa_{d,0}=m_{0}/M_{0}\approx 1$. This assumption does not affect the 
generality of the problem though higher values of $\kappa_{d,0}$ lead 
to a worse result from the point of view of convergence. 
The numerical nontrivial fixed point values are 
$\kappa_{d}^{*}\approx 0.34$ and $\tilde{g}^{*}\approx 21.4$.   
The numerical flow diagram in the 
$\tilde{g}\kappa_{d}$-plane is shown in Fig. 1. This result can be improved 
further by computing higher order terms and using resummation methods. 
Although we have shown the validity of the proposed scaling through 
perturbative methods, we claim that this scenario holds also beyond  
perturbation theory. As mentioned earlier, the present scaling with 
$e_{0,d}$ and $m_{0}/M_{0}$ fixed corresponds to freeze the 
amplitude fluctuations in the original GL model. Numerical 
simulations with frozen amplitudes but taking into account the phase 
fluctuations can be used as a final test of the above picture.

Note that the scaling we have just discussed predict the value 
$\nu'=1$ for the penetration depth exponent. This prediction implies 
the scaling $\lambda\sim\xi$ which is verified also in the vicinity of 
the inverted $XY_{3}$ critical point. It would be very interesting to 
verify this prediction experimentally and also by numerical means.    

Summarizing, we have considered two different scalings in the dGL model, 
each one corresponding to a different critical crossover. The first one 
gives the $XY_{3}$ scaling behavior which is probed experimentaly in 
single crystals of YBCO while the second one reproduces approximately 
the scaling behavior of BSCCO. Moreover, we predict the value 
$\nu'=1$ for the BSCCO penetration depth exponent.  
The results of this paper show the 
power of the dual approach. Indeed, all the possible crossovers 
obtained within the dGL model may be very difficult to obtain in the original 
GL model. The point is that a weak coupling regime in the dGL model corresponds 
to a strong coupling regime in the GL model. In this case nonperturbative 
effects become important.

Among the further directions is the inclusion of anisotropy and more general   
disorder parameters with higher symmetries like $d+is$ wave  
\cite{Heeb} and $SO(5)$ \cite{Zhang} symmetries. It would be also 
interesting to study other nouvel 
situations which occur in external field like 
the $\Phi$-transition of Te\u{s}anovi\'c \cite{Tesanovic}. Finally, we 
hope that the present contribution will stimulate some experimental 
effort towards the verification of the prediction $\nu'=1$.     

We would like to thank Z. Te\u{s}anovi\'c for bringing Ref. \cite{Junod} 
to our attention and A. Sudb{\o} for sending us Ref. \cite{Hove} prior 
publication. We would like to thank specially Prof. A. Junod for his 
valuable comments on the experimental results in YBCO and BSCCO.

\newpage

\begin{figure}
\centerline{\psfig{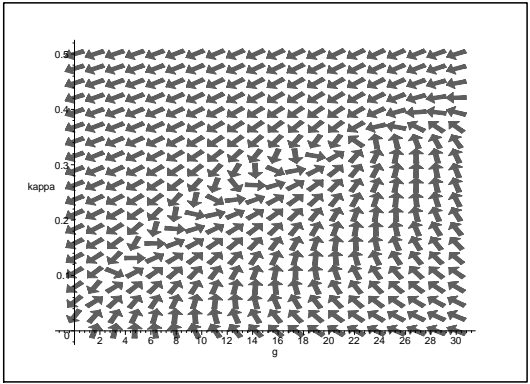}}
\caption{Flow diagram in the $\tilde{g}\kappa_{d}$-plane.}
\end{figure}


\begin{references}
\bibitem{Note} Another possibility is to take $m_{0}/M_{0}$ and 
$q_{0}$ fixed. In this case $e_{0,d}$ is not kept fixed as $M$ varies. 
Such a scaling has been considered before in Ref. \cite{Kiometzis1} using 
$m$ rather than $M$ as running scaling parameter. It leads to 
$\nu\approx 2/3$ and $\nu'=1/2$.
\end{references}
\end{document}